\documentclass[preprint,superscriptaddress,amsmath,amssymb,prd,aps,showpacs,floatfix,nofootinbib]{revtex4-1}
\usepackage{graphicx}
\usepackage{dcolumn}
\usepackage{bm}
\usepackage{mathrsfs}
\usepackage{hyperref}
\usepackage{amsmath}
\usepackage{amssymb}
\usepackage{amsfonts}
\usepackage{xcolor}
\usepackage{float}
\usepackage{dcolumn}
\usepackage{multirow}

\begin{document}
\title{Method to observe the $J^{P}=2^{+}$ partner of the $X_{0}(2866)$  in the $B^+ \to D^{+} D^{-} K^{+} $ reaction }

\date{\today}
\author{M.~Bayar}
\email{melahat.bayar@kocaeli.edu.tr}
\affiliation{Department of Physics, Kocaeli University, Izmit 41380, T{\"u}rkiye}
\affiliation{Departamento de
F\'{\i}sica Te\'orica and IFIC, Centro Mixto Universidad de
Valencia-CSIC Institutos de Investigaci\'on de Paterna, Aptdo.22085,
46071 Valencia, Spain}

\author{E.~Oset}
\email{oset@ific.uv.es}
\affiliation{Departamento de
F\'{\i}sica Te\'orica and IFIC, Centro Mixto Universidad de
Valencia-CSIC Institutos de Investigaci\'on de Paterna, Aptdo.22085,
46071 Valencia, Spain}

\begin{abstract}
We propose a method based on the moments of the $D^{-} K^{+}$ mass distribution in the $B^+ \to D^{+} D^{-} K^{+} $ decay to disentangle the contribution of the $ 2^{+} $ state, partner of $X_{0}(2900)$ in the $ \bar D^{*} K^{*}$  picture for this resonance. Some of these moments show the interference patterns of the $X_{1}(2900)$ and $X_{0}(2900)$ with the $ 2^{+} $ state, which provide a clearer signal of the $ 2^{+} $ resonance than the $ 2^{+} $ signal alone. The construction of these magnitudes from present data is easy to implement, and based on these data we show that clear signals for that resonance should be seen even with the present statistics. 
\end{abstract}

\maketitle


\section{Introduction}
The LHCb collaboration reported two states,  $X_{0}(2866)$ ($X_{0}(2900)$)  and $X_{1}(2900)$, in the $B^+ \to D^{+} D^{-} K^{+} $ decay from peaks in the $D^{-} K^{+}$ mass distribution \cite{LHCb:2020bls,LHCb:2020pxc}. The  states are manifestly exotic since they have two open quarks $\bar c \bar s $ and do not follow the standard  $q \bar q $ nature of mesons.  In the literature there are plenty of works using quark models about tetraquark states which could in principle accommodate such states \cite{rm1,rm2,rm3,rm4,rm5,rm8,rm9,rm7,rm6,rm11,rm10,hxchen,ahmed,pilloni,rosner,xliu,nora}. Yet, the proximity of the mass of the $X_{0}(2866)$  to the $ \bar D^{*} K^{*}$ threshold makes the molecular $ \bar D^{*} K^{*}$ picture appealing. In fact, ten years before its discovery, a state with $ I=0 $, $ J^{P}=0^{+} $ $  D^{*} \bar K^{*}$ molecule had been predicted with mass 2848 MeV and width around 23-59 MeV \cite{Molina:2010tx}, remarkably close to the experimental data of the $X_{0}(2866)$, $ M =2866 \pm 7$ MeV, $ \Gamma =57.2 \pm 12.9$ MeV. After the experimental discovery the molecular picture has been proposed in many work \cite{d14,d15,d16,d17,d18,d19}, but tetraquark pictures have also been advocated \cite{d5,d6,d7,d8}, one of them favoring the molecular structure \cite{Lu:2020qmp}. With the ordinary large uncertainties  in the mass, sum rules have also  contributed their share to the topic \cite{d9,d10,d11,d12,d13} and some of them support the molecular structure \cite{d11,d12,d13}. Other pictures have also been suggested, as peaks coming from analytical properties of triangle diagrams \cite{d22,d23,qifang}, or a triangular singularity \cite{d21}.

Coming back to the molecular picture, in \cite{Molina:2010tx}, together with the $ J^{P}=0^{+} $ state, two more bound states were predicted with $ J^{P}=1^{+}, 2^{+} $. The presence of the $ 0^{+}, 1^{+}, 2^{+} $ states is common in the studies of the vector vector interaction using the local hidden gauge approach \cite{Bando:1987br, Harada:2003jx, Meissner:1987ge, Nagahiro:2008cv} and extrapolations to the charm sector \cite{Molina:2010tx, Molina:2009ct}. Usually the $2^{+} $  state is the most bound, and actually the $ f_{2} (1270) $ as a $\rho\rho$ bound state \cite{Molina:2008jw, Geng:2008gx} is quite bound, to the point of being questioned as a molecular state in \cite{Gulmez:2016scm,Du:2018gyn}. However,  in \cite{Geng:2016pmf, Molina:2019rai} it was shown that the range of applicability of \cite{Gulmez:2016scm,Du:2018gyn} did not allow to make predictions in the very bound region. An improved method was proposed in \cite{Geng:2016pmf} corroborating the findings of \cite{Molina:2008jw, Geng:2008gx}. The picture is rather successful and in \cite{Geng:2008gx} one obtains the $ f_{2} (1270) $, $ f_{0} (1370) $, $ f'_{2} (1525) $, $ f_{0} (1710) $, $ K^{*}_{2} (1430) $ resonances, among others. The properties obtained give good explanations of radiative decays \cite{Nagahiro:2008um} and other decays of the resonances \cite{Oset:2012zza}. It is interesting mention that one of the predictions in \cite{Geng:2008gx} was a state of  $ I=1 $, partner of the $ f_{0} (1710)$. This state is also predicted in \cite{Du:2018gyn}, both of them with a mass around $1.75-1.79$ GeV. This state has been recently found by BaBar \cite{BaBar:2021fkz} and BESIII  \cite{BESIII:2021anf} collaborations and named $ a_{0} (1710) $. Using the information of \cite{Geng:2008gx}, a recent paper \cite{Dai:f0} shows the consistency of the prediction of   \cite{Geng:2008gx} with the findings of \cite{BaBar:2021fkz, BESIII:2021anf} and makes a prediction for the branching ratio of the  $ a_{0} (1710) $ production in the $D_{s}^+ \to \pi^{0} K^{+}  K_{s}^{0} $ reaction in agreement with the experimental rate obtained a posteriori in the BESIII experiment of Ref. \cite{BESIII:2022wkv}.

With this precedent of agreement of predictions with experiment one has confidence in the predictions done for the $  D^{*} \bar K^{*}$ states in \cite{Molina:2010tx}. Yet, the data about mass and width of the $X_{0}(2866)$  obtained \cite{LHCb:2020bls,LHCb:2020pxc} has served to fine tune the parameters of the theory (two form factors) of \cite{Molina:2010tx} in order to match exactly the mass and width of the  $X_{0}(2866)$ and this has been done in Ref. \cite{Molina:2020hde}. With the same parameters, predictions are made for the $ J^{P}=1^{+}, 2^{+} $ states and at the same time the decay widths for the  $  D^{*} \bar K^{*}$  ( $ 0^{+} $) state to $  D \bar K $, $  D^{*} \bar K^{*}$ ( $ 1^{+} $) to   $  D^{*} \bar K $ and $  D^{*} \bar K^{*}$  ( $ 2^{+} $) to   $  D \bar K $ and  $  D^{*} \bar K $ have been evaluated. The prediction for these states are shown in Table I.

\begin{table*}[tbh!]
\renewcommand\arraystretch{0.5}
\centering
\caption{$D^*\bar{K}^*$ states obtained by fine tuning of the free parameters including the width of the $D^*K$ channel.}\label{tab:tab1}
 \begin{tabular*}{1.00\textwidth}{@{\extracolsep{\fill}} c  c c c c}
 \hline
 \hline
  $I(J^P)$&$M[\mathrm{MeV}]$&$\Gamma[\mathrm{MeV}]$& Coupled channels & state\\
  \hline
  $0(2^+)$& $2775$ & $38$ & $D^*\bar{K}^*$ &?\\
  $0(1^+)$& $2861$ & $20$ & $D^*\bar{K}^*$ &?\\
  $0(0^+)$&$2866$ & $57$& $D^*\bar{K}^*$ &$X_0(2866)$\\
  \hline
  \hline
 \end{tabular*}
\end{table*}  

We shall use these data to make predictions with the method that we propose. The $B^+ \to D^{+} D^{-} K^{+} $  reaction can produce a $D^{-} K^{+}$ state with  $ J^{P}=0^{+} $ in $S$-wave. The  $D^{-} K^{+}$  in $ J^{P}=2^{+} $ requires $L=2$, which should reduce the strength of the amplitude and might explain why the $2^{+} $ state was not identified in the mass  $D^{-} K^{+}$  spectrum of this decay.  The $1^{+} $ state cannot be produced in this reaction for reasons of parity and angular momentum conservation. The  $1^{+} $ state still can be produced in $L=0$ in the $D^* \bar{K}$ spectrum of a suitable reaction, and two reactions were suggested to be used to search for this state: the $\bar{B}^0 \to D^{*+} \bar{D}^{*0} K^{-} $ reaction \cite{Dai:2022qwh}  and the  $\bar{B}^0 \to D^{*+} K^{-} K^{*0} $ \cite{Dai:2022htx}. Estimates were done for signals of the resonances and backgrounds, concluding that the signals should be seen clearly over the background coming from different sources. Yet, the  $ J^{P}=2^{+} $ could not be seen in similar reactions because it involves  $L=2$ and we cannot get related information from existing reactions. The idea of the present work is to propose an analysis method to be applied to the present or future $B^+ \to D^{+} D^{-} K^{+} $  data in order to extract the $2^{+} $ state so far not identified in these data \cite{LHCb:2020bls,LHCb:2020pxc}. The separation of $L=0$, $L=2$ amplitudes can in principle be done through a careful partial wave analysis of the data, where one anticipates problems if the $L=2$ amplitude is small. We propose a method to isolate the  $L=2$ part, and find very valuable interference patterns starting from the differential cross section $  \dfrac{d\Gamma}{dM_{inv}~ dcos \theta} $ without resorting to the partial wave analysis. It consists of using the momenta of this distribution making appropriate linear combinations of these magnitudes. We can isolate the  $L=0$ and $L=2$ part of the differential width $  \dfrac{d\Gamma}{dM_{inv}} $ and also produce an interference magnitude which can serve to identify the position of the $0^{+} $ and  $2^{+} $  resonances simultaneously. Since the interference magnitude is linear in the $2^{+} $ amplitude, the strength of the obtained pattern is bigger than the pure $2^{+} $  signal, proportional to the square of the $2^{+} $  amplitude,  which means that even with insufficient statistics to see the pure $2^{+} $ peak, an unequivocal signal of the existence of this resonance and its position could be found.

\section{Formalism}\label{formalism}

\begin{figure}[h!]
  \centering
  \includegraphics[width=0.90\textwidth]{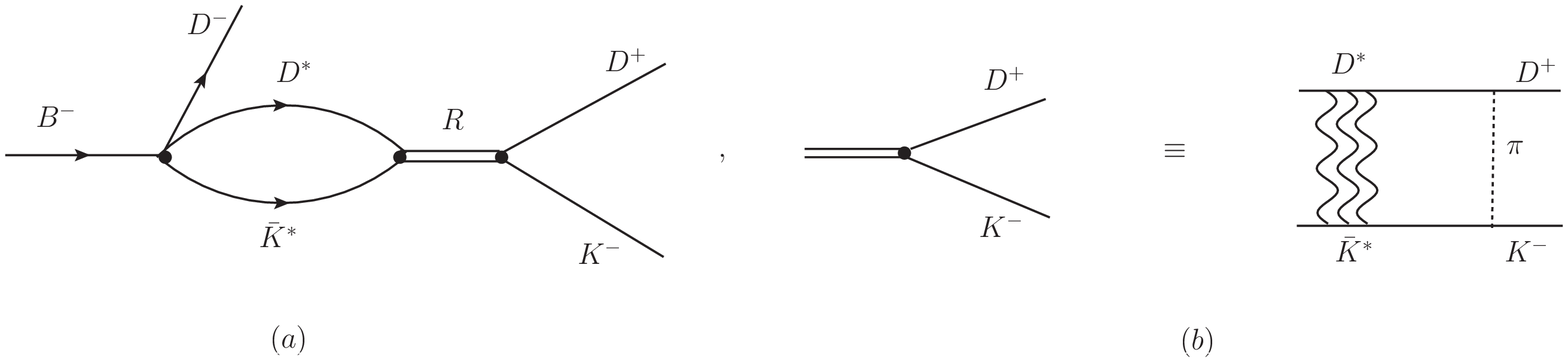}
  \caption{a) Diagrammatic mechanism for the $B^- \to D^{-} D^{+} K^{-} $ reaction mediated by the $0^{+} $,  $2^{+} $   $D^*\bar{K}^*$  resonances. b) Mechanism for the resonance decay into $D  \bar{K}$, mediated by $\pi$ exchange. The wiggly lines stand for sequential vector meson exchange. }
  \label{feynman}
\end{figure}

We shall use the $B^- \to D^{-} D^{+} K^{-} $ reaction to get information of the $D^*\bar{K}^*$ states of  \cite{Molina:2010tx, Molina:2020hde}. In order to understand how the reaction proceeds let us look at Fig.~\ref{feynman}. We produce in a first step $D^{-} D^*\bar{K}^* $, the $D^*\bar{K}^*$ couples to the resonance state and then it decays to $ D^{+} K^{-} $ in $ L=0 $ for $ J^{P}=0^{+} $ and in  $ L=2 $ for  $ J^{P}=2^{+} $. Note that the $1^{+} $ state cannot decays to $ D^{+} K^{-} $  since it would need $ L=1 $ and this violates parity. The $ P $-wave $D^* D \pi$ coupling in Fig.~\ref{feynman} (b) together with the also $ P $-wave $ K^* \to \bar{K} \pi $ coupling provide both $ L=0 $ and $ L=2 $ for the decays, which are studied in  \cite{Molina:2010tx, Molina:2020hde}. The coupling of the $D^*\bar{K}^*$  channel to the $J^{P}=0^{+},~ 2^{+}$ resonance $R$ occurs in $ S $-wave, the spin $J$ coming from the $D^*, ~\bar{K}^*$ spin combinations. If we produce the $0^+$ resonance then the $D^-$ will be in $ L=0 $ with respect to the resonance, but if we produce the $2^{+}$ resonance, the $D^-$ has to carry $ L=2 $ with respect to the resonance to match the zero angular momentum of the $B^-$. In order to see the structure of the amplitude for the mechanism of Fig.~\ref{feynman} (a) in the case of the $2^{+}$ state, we work in the rest frame of the $ D^{+} K^{-} $ system and recall that in terms of the polarization vectors of $D^*$ and $\bar{K}^*$  we have the spin projectors \cite{Molina:2008jw, Liang:2010ddf}

\begin{align}
P^{2}&=\frac{1}{2}(\epsilon_{i} \epsilon'_{j}+ \epsilon_{j} \epsilon'_{i})-\frac{1}{3} \epsilon_{l} \epsilon'_{l} \delta_{ij} \label{eq:P2P0}  \\ 
P^{0}&=\frac{1}{3} \epsilon_{l} \epsilon'_{l} \delta_{ij} \nonumber
\end{align}
with $ \vec{\epsilon}$, $ \vec{\epsilon}~'$ applying to the  $D^*$ and  $\bar{K}^*$ polarization vectors respectively. By calling $ \vec{p}$ the momentum of the $D^-$ and $ \vec{q}$  the momentum of the $K^{-} $ in that frame, we have the vertices

\begin{equation}
\left( p_{i} p_{j} -\frac{1}{3} \vec{p}^{~2} \delta_{ij} \right) \left[ \frac{1}{2}(\epsilon_{i} \epsilon'_{j}+ \epsilon_{j} \epsilon'_{i})-\frac{1}{3} \epsilon_{l} \epsilon'_{l} \delta_{ij} \right]  \label{eq:vertex}  
\end{equation}
for the  $D^*\bar{K}^*$ of the resonance coupling to the $D^-$ in $ D $-wave, and 

\begin{equation}
 \left[ \frac{1}{2}(\epsilon_{i'} \epsilon'_{j'}+ \epsilon_{j'} \epsilon'_{i'})-\frac{1}{3} \epsilon_{l'} \epsilon'_{l'} \delta_{i'j'} \right]  \left( q_{i'} q_{j'} -\frac{1}{3} \vec{q}^{~2} \delta_{i'j'} \right)  \label{eq:vertexD}  
\end{equation} 
for the coupling of the resonance to the $ D^{+} K^{-} $ final component in  $ D $-wave. By summing over the polarizations of the polarization vectors the product of the vertices of Eqs. (\ref{eq:vertex}), (\ref{eq:vertexD}), and considering that 

\begin{equation}
\sum_{pol}{ \epsilon_{i} \epsilon_{j}} = \delta_{ij}  \label{eq:sumpol} ,
\end{equation}
we obtain the magnitude
\begin{equation}
(\vec{p}.\vec{q})^{~2}-\frac{1}{3} \vec{p}^{~2} \vec{q}^{~2} = \vec{p}^{~2} \vec{q}^{~2} \left( cos^{2} \theta -\frac{1}{3} \right) =
 \frac{2}{3} \sqrt{\frac{ 4 \pi}{5} } \vec{p}^{~2} \vec{q}^{~2} Y_{20}(cos \theta ) \label{eq:sumpol} ,
\end{equation}
where $  \theta $ is the angle between the $ K^- $  and $D^-$  momenta in the  $ D^{+} K^{-} $ rest frame. Let us note that if we make a boost from the  $ B^- $ rest frame to the $ D^{+} K^{-} $ rest frame,  the $D^-$ direction does not change and hence, this angle can equally be considered the one between the $ K^- $  in the  $ D^{+} K^{-} $ rest frame and the one of the $D^-$ in the $B^-$ rest frame. This is worth mentioning because these variables are those usually considered in the evaluation of the differential cross section, which is given by

\begin{eqnarray}
\frac{d\Gamma}{dM_{\rm inv} d \tilde{\Omega}}=\frac{1}{(2\pi)^4} \frac{1}{8 M^{2}_{B}} ~p_{D^-} ~\tilde{k} ~|t|^2 \label{eq:GdMinvdotil}
\end{eqnarray}
where $dM_{\rm inv} $ is the $ D^{+} K^{-} $ invariant mass, $t $ the transition matrix, and $p_{D^-}$, $ \tilde{k} $ the momenta of the $D^-$ and $ K^{-} $ in the $ B^- $ rest frame and $ D^{+} K^{-} $ rest frame, respectively

\begin{equation}
 p_{D^-}=\frac{\lambda^{1/2}(M^2_{B},m^2_{D},M^2_{\rm inv})}{2 M_{B}} \nonumber
 \end{equation}

\begin{equation}
  \tilde{k}\equiv q=\frac{\lambda^{1/2}(M^2_{\rm inv},m^2_{D},m^2_{K})}{2 M_{\rm inv}} \nonumber
\end{equation}
and $\tilde{\Omega}$ the solid angle in the $ D^{+} K^{-} $ rest frame.

\begin{figure}[h!]
  \centering
  \includegraphics[width=0.50\textwidth]{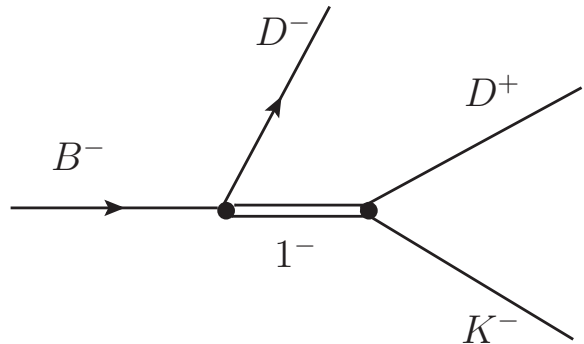}
  \caption { Decay mechanism for $B^- \to D^{-} D^{+} K^{-} $ through a $ 1^- $  state coupling to $ D^{+} K^{-} $. }
  \label{feynman2}
\end{figure}

The production of the $ 0^+ $ resonance has no angular structure since it occurs in $S$-wave (one can do the same exercise before with the $ \epsilon_{l} \epsilon'_{l} $ structure). Since we can only have $ 0^+ $ and $ 2^+ $ resonance contribution with positive parity in the amplitude, since the $ 1^+ $ was excluded for parity  reasons, our full resonant amplitude will have the contribution of the $ 0^+ $ and $ 2^+ $  states (see Fig.~\ref{feynman} (a)), but in addition one can also have the $ 1^- $ contribution. Indeed, in the LHCb experiment a $ 1^- $ state,  $X_{1}(2900)$, was also reported, and, according to Ref. \cite{LHCb:2020pxc} it has a bigger strength in the $ D^{-} K^{+} $ distribution ($ D^{+} K^{-} $ for us) than the  $X_{0}(2900)$. The $ 1^- $ state contribution is depicted in Fig.~\ref{feynman2}

In this mechanism the $B^- \to D^{-} R(1^-) $ vertex requires a $P$-wave, and the decay of the $ 1^- $ resonance to $ D^{+} K^{-} $ also involves a $P$-wave. Using the same argument as before, the amplitude will now have the structure 

\begin{equation}
\vec{p}.\vec{q}= p~q \sqrt{\frac{4 \pi}{3} }  Y_{10} \label{eq:structure1} . 
\end{equation}
Adding the three resonance contributions in the amplitude we have the structure 

\begin{eqnarray}
t &=& \alpha~ G_{D^* \bar{K}^*} (M_{\rm inv}) \dfrac{1}{M_{\rm inv}^{2} -M^{2}_{X_{0}} +i~M_{X_{0}}~\Gamma_{X_{0}}} Y_{00}(cos \theta ) \nonumber\\
&+& \beta ~G_{D^* \bar{K}^*}  (M_{\rm inv}) \dfrac{1}{M_{\rm inv}^{2} -M^{2}_{X_{2}} +i~M_{X_{2}}~\Gamma_{X_{2}}} Y_{20}(cos \theta )\nonumber\\ 
&+& \gamma ~ \dfrac{1}{M_{\rm inv}^{2} -M^{2}_{X_{1}} +i~M_{X_{1}}~\Gamma_{X_{1}}} Y_{10}(cos \theta ) \nonumber\\ 
&\equiv & a ~Y_{00} (cos \theta )+b ~Y_{20} (cos \theta )+c ~Y_{10}(cos \theta ) \label{eq:tmat}
\end{eqnarray}
with 
\begin{eqnarray}
a &=& \alpha~ G_{D^* \bar{K}^*} (M_{\rm inv}) \dfrac{1}{M_{\rm inv}^{2} -M^{2}_{X_{0}} +i~M_{X_{0}}~\Gamma_{X_{0}}} \nonumber\\
b&=& \beta ~G_{D^* \bar{K}^*}  (M_{\rm inv}) \dfrac{1}{M_{\rm inv}^{2} -M^{2}_{X_{2}} +i~M_{X_{2}}~\Gamma_{X_{2}}} \nonumber\\ 
c&=& \gamma ~ \dfrac{1}{M_{\rm inv}^{2} -M^{2}_{X_{1}} +i~M_{X_{1}}~\Gamma_{X_{1}}}  \nonumber\\ 
 \label{eq:abc}
\end{eqnarray}
where $  G_{D^* \bar{K}^*} (M_{\rm inv}) $ is the loop function of $D^*\bar{K}^*$ , which we take from Ref. \cite{Molina:2020hde}, and the masses and widths are taken from the experiment for $ X_{0} $, $ X_{1} $ and from Table \ref{tab:tab1} for $ X_{2} $.

From Eq. (\ref{eq:tmat})  we find that 

\begin{eqnarray}
\vert t \vert^{2} &= & \vert a \vert^{2} ~Y_{00}^{2}+\vert b \vert^{2} ~Y_{20}^{2}+\vert c \vert^{2} ~Y_{10}^{2}+2~ Re(ab^{*})~Y_{00}~Y_{20}\nonumber\\
&+&2~ Re(ac^{*})~Y_{00}~Y_{10}+2~ Re(bc^{*})~Y_{20}~Y_{10} \label{eq:tmatsq}
\end{eqnarray}

Then we define the following magnitudes

\begin{eqnarray}
\frac{d~\Gamma_{i}}{dM_{\rm inv} }=  \int d \tilde{\Omega}~ \dfrac{d~\Gamma}{dM_{\rm inv} ~d \tilde{\Omega}}~Y_{i0}  \label{eq:dGDMinv}
\end{eqnarray}
for $i=0,1,2,3,4$. 

These magnitudes are a simplified version of the moments of the distribution, where the projections are done with the functions $d_{mm'}^{L}$. Recent examples of application of these moments can be seen in \cite{miguelguo} studying heavy light meson spectroscopy, in   \cite{Migueladam} calculating moments for $\eta \pi^{0}$ photoproduction at GlueX, and in analysis by the LHCb collaboration of the $B^- \to  D^{+} \pi^{-} \pi^{+}$ \cite{LHCbnew} and $B_{s}^0 \to  \bar{D}^{0} K^{-} \pi^{+}$ \cite{LHCbnewdos} reactions. Considering that 

\begin{equation}
d_{00}^{L }=\left( \dfrac{4 \pi}{2L+1} \right)^{1/2}  Y_{L0}(cos \theta ) \nonumber \label{eq:d00L} ,
\end{equation}
our results should agree with those of standard moments multiplying the latter ones by $((2L+1)/4 \pi)^{1/2}$, but we   rederive  them here. By using orthogonality relations of the spherical harmonics and the integrals of three spherical harmonics   \cite{rose}, we easily find the following formulas 

\begin{equation}
\dfrac{d~\Gamma_{0}}{dM_{\rm inv} }=FAC \left[ \vert a \vert^{2} +\vert b \vert^{2} +\vert c \vert^{2} \right]  \label{eq:dG0dMinv} ,
\end{equation}

\begin{equation}
\dfrac{d~\Gamma_{1}}{dM_{\rm inv} }=FAC \left[ 2~ Re(ac^{*}) + \frac{2}{\sqrt{5}} ~2~ Re(bc^{*}) \right]  \label{eq:dG1dMinv} ,
\end{equation}

\begin{equation}
\dfrac{d~\Gamma_{2}}{dM_{\rm inv} }=FAC \left[ \frac{2}{7}\sqrt{5} ~\vert b \vert^{2} + \frac{2}{5}\sqrt{5} ~\vert c \vert^{2} +2~ Re(ab^{*})\right]  \label{eq:dG2dMinv} ,
\end{equation}

\begin{equation}
\dfrac{d~\Gamma_{3}}{dM_{\rm inv} }=FAC  \sqrt{\frac{15}{7}} ~\frac{3}{5} ~2~ Re(bc^{*})  \label{eq:dG3dMinv} ,
\end{equation}

\begin{equation}
\dfrac{d~\Gamma_{4}}{dM_{\rm inv} }=FAC ~\frac{6}{7} \vert b \vert^{2} \label{eq:dG4dMinv} ,
\end{equation}
with
\begin{equation}
FAC =\frac{1}{\sqrt{4 \pi}} \frac{1}{(2 \pi)^4} \frac{1}{8 M^{2}_{B}} ~ p_{D^-}~ \tilde{k} \nonumber  \label{eq:fac}. 
\end{equation}

One can see that indeed these formulas agree with the momenta of Eqs. (2)-(6) of \cite{LHCbnew}  considering  $ 0^+, 1^-, 2^+ $ resonances, upon the correction factor $((2L+1)/4 \pi)^{1/2}$ discussed above. The derivation using $Y_{L0}(cos \theta )$, which is what we need here, is very easy to do as we have explained above. 

The magnitudes of Eqs. (\ref{eq:dG0dMinv})-(\ref{eq:dG4dMinv}) can be easily constructed from the experimental distribution $\dfrac{d~\Gamma}{dM_{\rm inv} d \tilde{\Omega}}$, without the need of making a partial wave analysis, and are very instructive. First we realize that  $d\Gamma_{4}/dM_{\rm inv}$ filters the $2^+$ contribution. This is very intuitive since the magnitude $d\Gamma_{4}/dM_{\rm inv}$ has eliminated the contribution of the $ 0^+$ and $ 1^- $ states and should show a clear peak for the $2^+ $ state, provided one has enough statistics. The magnitude  $d\Gamma_{3}/dM_{\rm inv}$ has only one term and measures interference between the  $2^+$  and  $ 1^- $ resonances. Given the fact that this magnitude is linear in the $2^+$ amplitude (the $b$ factor) its strength should be bigger than $d\Gamma_{4}/dM_{\rm inv}$, which is  proportional to $ \vert b \vert^{2} $, and could help see the  $2^+$ structure which is not clearly seen in the experimental $d\Gamma/dM_{\rm inv}$ spectrum where the $2^+$ contribution also goes $ \vert b \vert^{2} $. 

There are other relations that we can obtained by linear combination of  $d\Gamma_{i}/dM_{\rm inv}$. Eq. (\ref{eq:dG3dMinv}) can be substituted in Eq. (\ref{eq:dG1dMinv}) to obtain $ Re(ac^{*}) $ and using Eqs. (\ref{eq:dG4dMinv}) in Eqs. (\ref{eq:dG0dMinv}) and (\ref{eq:dG2dMinv}) we can also eliminate the $ \vert c \vert^{2} $ term. Then we get two more equations and altogether we find:

\begin{equation}
FAC ~ \vert b \vert^{2}  = \frac{7}{6}  \dfrac{d~\Gamma_{4}}{dM_{\rm inv} }  \label{eq:facb2} ,
\end{equation} 

\begin{equation}
FAC ~ 2~ Re(bc^{*}) = \frac{5}{3} \sqrt{\frac{7}{15}}  \dfrac{d~\Gamma_{3}}{dM_{\rm inv} }  \label{eq:intbcstr} ,
\end{equation} 

\begin{equation}
FAC ~2~ Re(ac^{*}) = \dfrac{d~\Gamma_{1}}{dM_{\rm inv} } - \frac{2}{3} \sqrt{\frac{7}{3}}  \dfrac{d~\Gamma_{3}}{dM_{\rm inv} }  \label{eq:intacstr} ,
\end{equation} 

\begin{equation}
FAC \left[\vert a \vert^{2} - \frac{\sqrt{5}}{2} ~2~ Re(ab^{*}) \right] = \dfrac{d~\Gamma_{0}}{dM_{\rm inv} }  -\frac{\sqrt{5}}{2}  \dfrac{d~\Gamma_{2}}{dM_{\rm inv} } - \frac{1}{3} \dfrac{d~\Gamma_{4}}{dM_{\rm inv} } \label{eq:a2intacstr} .
\end{equation}

We can see that Eq. (\ref{eq:facb2}) provides the net contribution of the  $2^+$ state to the mass distribution.  Eq. (\ref{eq:intbcstr}) provides an interference of the $2^+$ and $1^-$ resonances. Eq. (\ref{eq:intacstr}) the interference of the $ 0^+$ and $ 1^- $ resonances and Eq. (\ref{eq:a2intacstr}) gives the contribution of the $ 0^+$ together with an interference term from the $ 0^+$ and $2^+$  resonances.  In the next section we see what we expect from these magnitudes using information already contained in the experimental analysis of Ref. \cite{LHCb:2020pxc}.

\section{Results}\label{result}

We rely upon data from the analysis of Ref. \cite{LHCb:2020pxc} that led to the claims of the  $X_{0}(2900)$ and $X_{1}(2900)$ resonances. We use these data to choose the parameters $ \alpha, \gamma $ of the amplitude of Eq. (\ref{eq:tmat}). The parameter $ \beta $ which determines the strength of the $2^+$ state is chosen to be consistent with the data of \cite{LHCb:2020pxc} and its not obvious observation in this experiment. With reasonable numbers for $ \beta $ consistent with the experimental data we evaluate the magnitudes discussed in the former section. 

\begin{figure}[h!]
  \centering
  \includegraphics[width=0.80\textwidth]{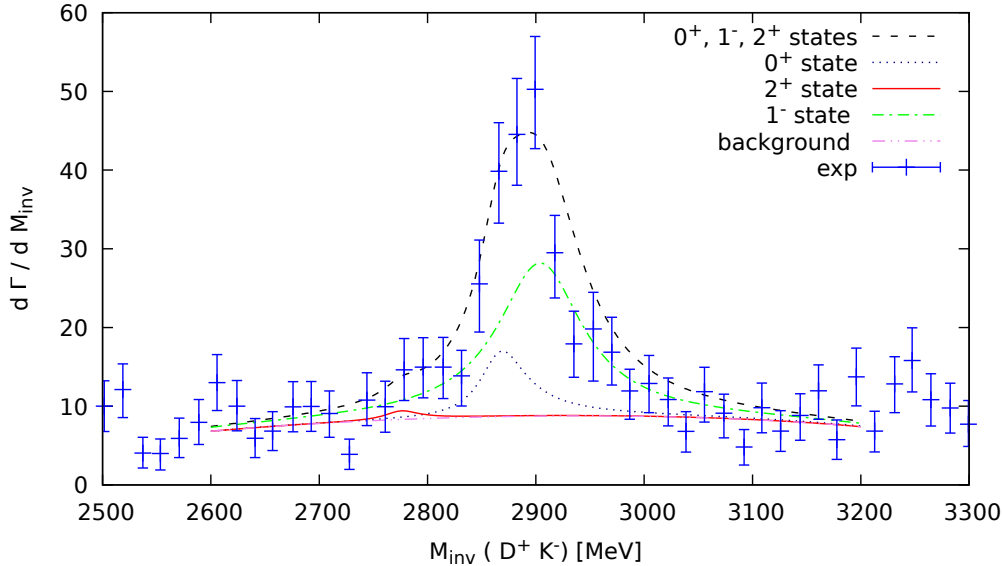}
  \caption{(color online) $ \dfrac{d~\Gamma_{0}}{dM_{\rm inv} } $.  Contribution of $ 0^+$ , $ 1^- $ and $2^+$  resonances (black dashed line), $ 0^+$ (navy dotted line) , $ 1^- $  (green dashed-dotted line) and $2^+$ (solid red line)  states alone with background (violet dashed double dotted line) for the process  $B^+ \to D^{+} D^{-} K^{+} $.}
  \label{figall}
\end{figure}

In Fig. \ref{figall} we plot $ \dfrac{d~\Gamma_{0}}{dM_{\rm inv} } $ of Eq. (\ref{eq:dG0dMinv}) and fix the parameters  $ \alpha, \gamma $ such as to approximately give the $X_{0}$, $X_{1}$ signals as shown in Fig. 12 of \cite{LHCb:2020pxc}. We also add a small background to approximately reproduce the one in \cite{LHCb:2020pxc}, not tied to the $X_{0}$, $X_{1}$ resonances, which simply follows phase space, easily obtained taking $ \vert t \vert^{2} $ equal to a constant in Eq. (\ref{eq:GdMinvdotil}). We adjust the parameter $ \beta $ such as to give some contribution to the mass distribution in the region of $ 2775 $ MeV which is consistent with a small but clear bump in the data. The agreement with the data with this set up is fair, such that we can proceed to plot the other magnitudes.

\begin{figure}[h!]
  \centering
  \includegraphics[width=0.80\textwidth]{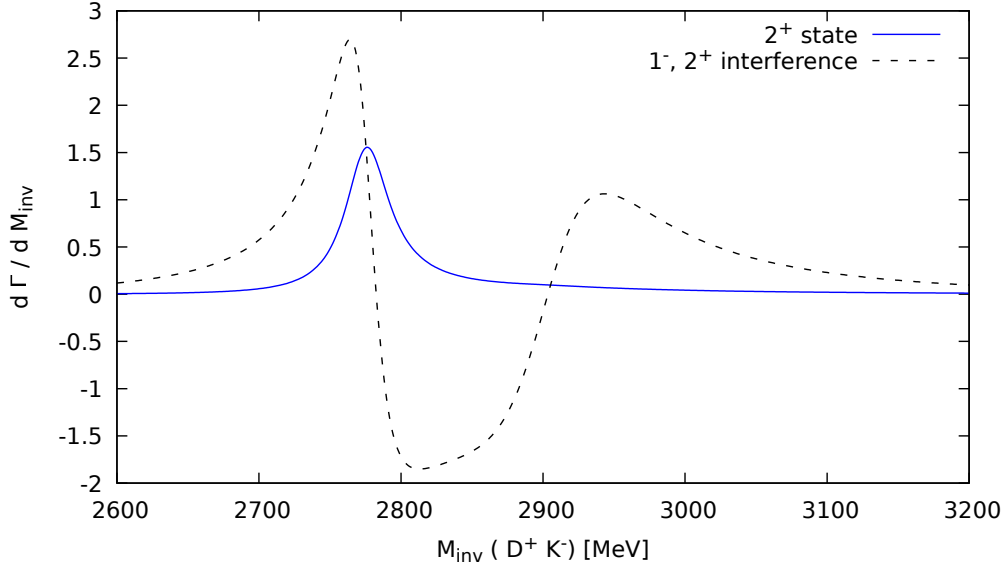}
  \caption{(color online) $ \dfrac{7}{6}  \dfrac{d~\Gamma_{4}}{dM_{\rm inv} } $ ($2^+$ state alone; solid blue line) and  $  \dfrac{5}{3} \sqrt{\dfrac{7}{15}}  \dfrac{d~\Gamma_{3}}{dM_{\rm inv} } $ (the interference of the $X_{1}$ with the  $X_{2}$ resonances; black dashed line) for the process  $B^+ \to D^{+} D^{-} K^{+} $.}
  \label{Cont2pandX1X2int}
\end{figure}

In Fig. \ref{Cont2pandX1X2int} we plot the magnitudes $ \dfrac{7}{6}  \dfrac{d~\Gamma_{4}}{dM_{\rm inv} }  $ and $  \dfrac{5}{3} \sqrt{\dfrac{7}{15}}  \dfrac{d~\Gamma_{3}}{dM_{\rm inv} } $, which provide the contribution of the  $2^+$ state alone, and the interference of the $X_{1}$ with the  $X_{2}$ resonances respectively. We have eliminated the possible contribution of the background to these magnitudes. It would not contribute to the structure seen for the  $ 1^- $,  $2^+$ interference around the  $2^+$ position if it has no $ 1^- $ component, and a possible small constant contribution in this channel still would produce a similar shape. What we observe there is interesting. Since  $  \dfrac{d~\Gamma_{4}}{dM_{\rm inv} }  $ singles out the  $2^+$ contribution, the signal stands very clear, and hence, this magnitude, easily obtainable from the experimental $D^{+} K^{-}$ mass distribution, should serve to show if there is a $2^+$ state. Since the process proceeds via a $ D $- wave, we anticipate a strength smaller than for the $X_{0}$ or $X_{1}$. But the fact that this magnitude eliminates any contribution of $X_{0}$, $X_{1}$, or interferences of any of the three resonances, makes it ideal to identify a likely $2^+$ state. The information obtained from $ \dfrac{d~\Gamma_{3}}{dM_{\rm inv} } $ is no less important because it shows a different property of the resonance, the approximate behavior of the real part of the amplitude going through zero at the resonance peak. We find this structure for the interference of the $X_{1}$ and  $X_{2}$ resonances. Since the signal is proportional to $ b $ rather than $ \vert b \vert^{2} $ (a small magnitude), the strength of the signal is magnified with respect to the one of the $2^+$  contribution alone. This information is additional to the one obtained from the $2^+$ contribution alone, and the observation of the two magnitudes, even with the present statistic should give us a clear answer about the existence of the $2^+$ resonance and its properties. We should note that $ Re(bc^{*}) $ is proportional to 
\begin{equation}
(M^{2}_{\rm inv}-M^{2}_{X_{2}})(M^{2}_{\rm inv}-M^{2}_{X_{1}})+M_{X_{1}} \Gamma_{X_{1}} M_{X_{2}} \Gamma_{X_{2}} 
 \label{eq:compbc}
\end{equation}
and we can see that, assuming the second term above small, is goes through zero for $ M_{\rm inv} =M_{X_{2}}$ with negative slope in $M_{\rm inv}$, and also goes through zero for $ M_{\rm inv} =M_{X_{1}}$ with positive slope as is the case in the figure. Also, the different masses of  $X_{1}$ and  $X_{2}$, make big the first term, which is responsible for this structure. The same formula applied to  $X_{0}$ and $X_{1}$ which are very close, blurs this pattern and the signals of resonances are not so clear, as we show below.

\begin{figure}[h!]
  \centering
  \includegraphics[width=0.80\textwidth]{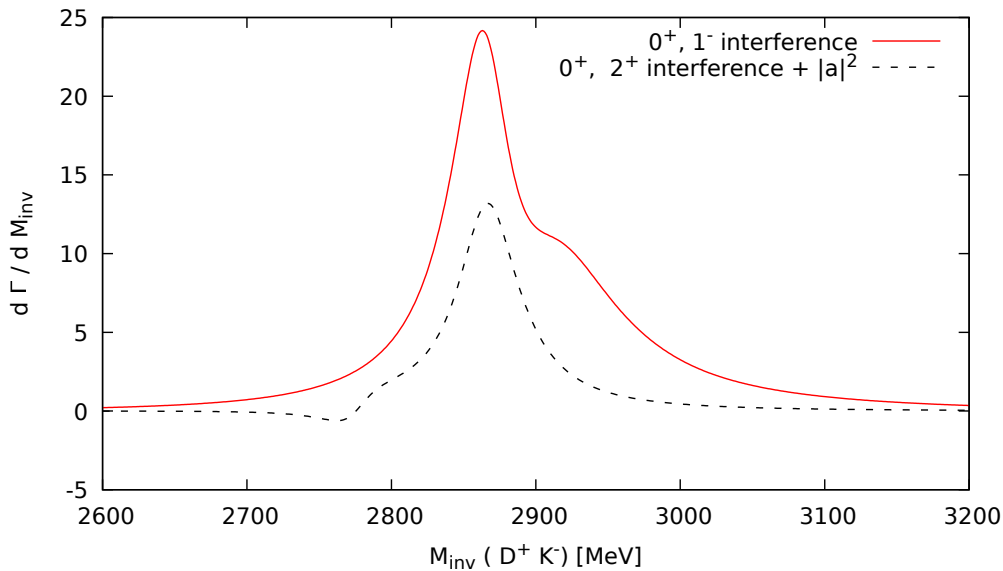}
  \caption{(color online) $ \dfrac{d~\Gamma_{1}}{dM_{\rm inv} } - \dfrac{2}{3} \sqrt{\dfrac{7}{3}}  \dfrac{d~\Gamma_{3}}{dM_{\rm inv} }  $ (the interference of the $X_{0}$ and $X_{1}$ resonances; solid red line) and  $ \dfrac{d~\Gamma_{0}}{dM_{\rm inv} }  -\dfrac{\sqrt{5}}{2}  \dfrac{d~\Gamma_{2}}{dM_{\rm inv} } - \dfrac{1}{3} \dfrac{d~\Gamma_{4}}{dM_{\rm inv} } $ (a mixture of the $0^+$ contribution and the interference of  $0^+$ and $2^+$; black dashed line) for the process  $B^+ \to D^{+} D^{-} K^{+} $. }
  \label{intX0X1}
\end{figure}

In Fig. \ref{intX0X1} we show another plot with the magnitudes of Eqs. (\ref{eq:intacstr}), (\ref{eq:a2intacstr}). The first magnitude measures the interference of the $X_{0}$ and $X_{1}$ resonances. It gives no information on the  $2^+$, and, as anticipated, the interference of  $X_{0}$, $X_{1}$ has not a very clear pattern. Finally the magnitude of Eq. (\ref{eq:a2intacstr}) provides a mixture of the $0^+$ contribution and the interference of  $0^+$ and $2^+$. It shows a similar pattern as the interference of $1^-$ and $2^+$ but the strength of the interference is smaller than for the case of  $1^-$ and $2^+$, because the strength of the $0^+$ is smaller than that of the $1^-$ .

\section{Conclusions}\label{conclusion}

We have presented here a method based on the moments of the mass distribution that we find particularly suited to isolate contributions of different resonances present in the $  D^{+} K^{-} $ mass distribution of a $ B^{-} $ decay to $ D^{-} D^{+} K^{-} $. Our purpose was to identify the presence of a $2^+$ contribution in the data. Assuming that it is small, since otherwise the state would have already been claimed, we find that the magnitude obtained projecting the mass distribution (not the amplitude) with the $ Y_{40} $ spherical harmonic singles out the $2^+$ contribution to the mass distribution with about the same weight as it comes in the global mass distribution. Since one does not have to make a partial wave analysis including many contributions and summing them coherently in the amplitude, the process suggested is certainly most welcome.

We also show that the projection of the mass distribution over the spherical harmonic $ Y_{30} $ provides an interference pattern of the $2^+$ and $1^-$ contribution which allows one to identify the two resonances clearly. 

We also provide other interesting magnitudes by projecting over the spherical harmonics $ Y_{10} $, $ Y_{20} $, and making appropriate linear combinations of these magnitudes that offer information on the interference of $0^+$ and $1^-$ resonances and of the  $0^+$ and $2^+$. 

The magnitudes suggested are easy to implement from the experimental data of $\dfrac{d~\Gamma}{dM_{\rm inv} d \tilde{\Omega}}$ and they should provide an important complement to the standard partial wave analysis of Ref. \cite{LHCb:2020pxc}, from where we are reasonably confident that evidence for the $2^+$ state, partner of the  $X_{0}(2900)$ in the $ D^{*} \bar{K}^{*} $  molecular picture should emerge. The results obtained in this paper should definitely encourage this easy analysis.

\section{ACKNOWLEDGEMENT}
This work is partly supported by the Spanish Ministerio
de Economia y Competitividad (MINECO) and European FEDER funds under Contract No. PID2020-112777GB-I00, and
by Generalitat Valenciana under contract PROMETEO/2020/023. This project has received funding from the European Union
Horizon 2020 research and innovation programme under the program H2020-INFRAIA-2018-1, grant agreement No. 824093
of the STRONG-2020 project.


\end{document}